# 3D time-resolved analysis of the evolution metamagnetic phase transition in FeRh system

Aleksei S. Komlev[a]*, Rodion A. Makarin[a], Tatiana S. Ilina[d], Dmitry A. Kiselev[d], Alisa M. Chirkova[b], Nikita A. Kulesh[c], Alexey S. Volegov[c], Vladimir I. Zverev[a], Nikolai S. Perov[a]

[a] *Faculty of Physics, M.V. Lomonosov Moscow State University, 119991, Moscow, Russia.*
[b] *TU Darmstadt, Institute for Materials Science, 64287, Darmstadt, Germany*
[c] *Institute of Natural Sciences, Ural Federal University, 620083, Yekaterinburg, Russia*
[d] *National University of Science and Technology "MISiS", 119049, Moscow, Russia*

**Abstract**

The FeRh alloy is an attractive material for studies of magnetic first-order phase transitions. The phase transition in FeRh from an antiferromagnetic to the ferromagnetic state is accompanied by the nucleation, growth, and merging of ferromagnetic clusters. The ferromagnetic phase evolution is studied in detail, and its various stages are distinguished. Both static properties and phase transition kinetics (time dependences of magnetization) are investigated. A comprehensive analysis allows us to determine the ferromagnetic phase nucleation regions and the main features of phase growth. In addition, the mechanisms that lead to a change in the ferromagnetic phase growth rate (the limitation effect and the magnetocaloric effect) were determined. The variation of phase transition evolution dominant mechanisms depending on the sample microstructure was shown.

*Corresponding author. Email address: komlev.as16@physics.msu.ru

# Introduction

Materials that exhibit a first-order magnetostructural phase transition often have transition mechanisms that are difficult to interpret. The main reason is the interdependence of the magnetic, structural and electronic properties of such materials. Therefore, it can be complicated to establish the driving forces of a metamagnetic transition. On the other hand, the presence of a correlation between various physical properties of such materials allows to observe a number of anomalous effects. For example, there is a giant magnetocaloric effect [1,2], colossal magnetoresistance [3], giant magnetostriction [4,5], anomalous thermoelectric effects [6,7] near the temperature of the magnetic phase transition. These effects have a wide range of promising practical applications. In particular, it was proposed to create environmentally friendly refrigeration devices based on the magnetocaloric effect [8]. Improvements of technologies for the synthesis of new materials with giant effects open the way for new practical applications of magnetocaloric materials. There are several successful attempts to integrate these materials in spintronics [6,9], energy harvesting [10], medicine [11,12], and in information recording devices with the HAMR (heat-assisted magnetic recording) system [13]. The problem of the origin of the above-mentioned effects remains still unresolved. The main obstacle is the lack of an *ab initio* description of the first order magnetic phase transition mechanisms.

The binary FeRh alloy of near-equiatomic stoichiometric composition is fascinating model-material for first order magnetic phase transition analysis. The antiferromagnetic-ferromagnetic phase transition occurs near room temperature. The cubic symmetry of the crystal lattice does not change during the transition. This alloy demonstrates record values of a number of effects [1,6,8]. The listed factors simplify the process of obtaining experimental results and their correct interpretation for FeRh alloys. The rearrangement of the crystalline and magnetic substructures in this alloy occurs simultaneously according to previous theoretical [14] and experimental [15] studies. Therefore, it is believed that the change in the electronic structure is the dominant mechanism of the phase formation . It is worth noting that the static physical properties of this compound have been well studied [17,18]. It is known that tiny variations in the elemental composition may change the phase transition temperature [18], as well as the appearance of tensile leads to an increase in the phase transition temperature [19,20]. Mechanical stresses may arise inside the bulk sample due to the presence of an additional γ-phase [21]. Stresses in film samples are created by the substrate [20,22,23]. It was found that the new phase nucleation occurs, as a rule, on sample defects using such methods as MFM [24,25], XMCD [26], MOKE [27], Lorentz microscopy [28,29]. It must be mentionedthat the dynamics of the ferromagnetic phase nucleation processes remains not fully investigated. The results of MOKE [30–32], XPS [16], XRD [33,34], XMCD [35] measurements performed using the pump-probe method are now available. These measurements demonstrate the ultrafast dynamics of the system after a pump pulse (during ten nanoseconds). In the meantime, magnetization measurements by VSM [36–39] and XMCD [40] methods over a longer time interval demonstrate magnetization relaxation within several minutes. Such long relaxation times cannot be explained in terms of the Landau-Lifshitz equation applicability [41–43]. It should also be taken into account that the nature of the phase transition is determined not only by the physical parameters of the material, but also by the microstructure of the sample [44–46]. Therefore, an additional analysis of the ferromagnetic phase evolution processes is required.

## Main

This article presents the results of a study of bulk alloys with $Fe_{49}Rh_{51}$ and $Fe_{50}Rh_{50}$ compositions and a 56 nm thick film ($Fe_{49}Rh_{51}$) grown on a single-crystal MgO substrate. The bulk $Fe_{49}Rh_{51}$ samples have different volume fractions of the paramagnetic γ-phase (5% and 35%). All other samples lack of the γ-phase. All samples were annealed in an argon atmosphere. More information on synthesis and sample preparation can be found in **Supplementary materials** and **Methods**.

**Magnetic characterization**

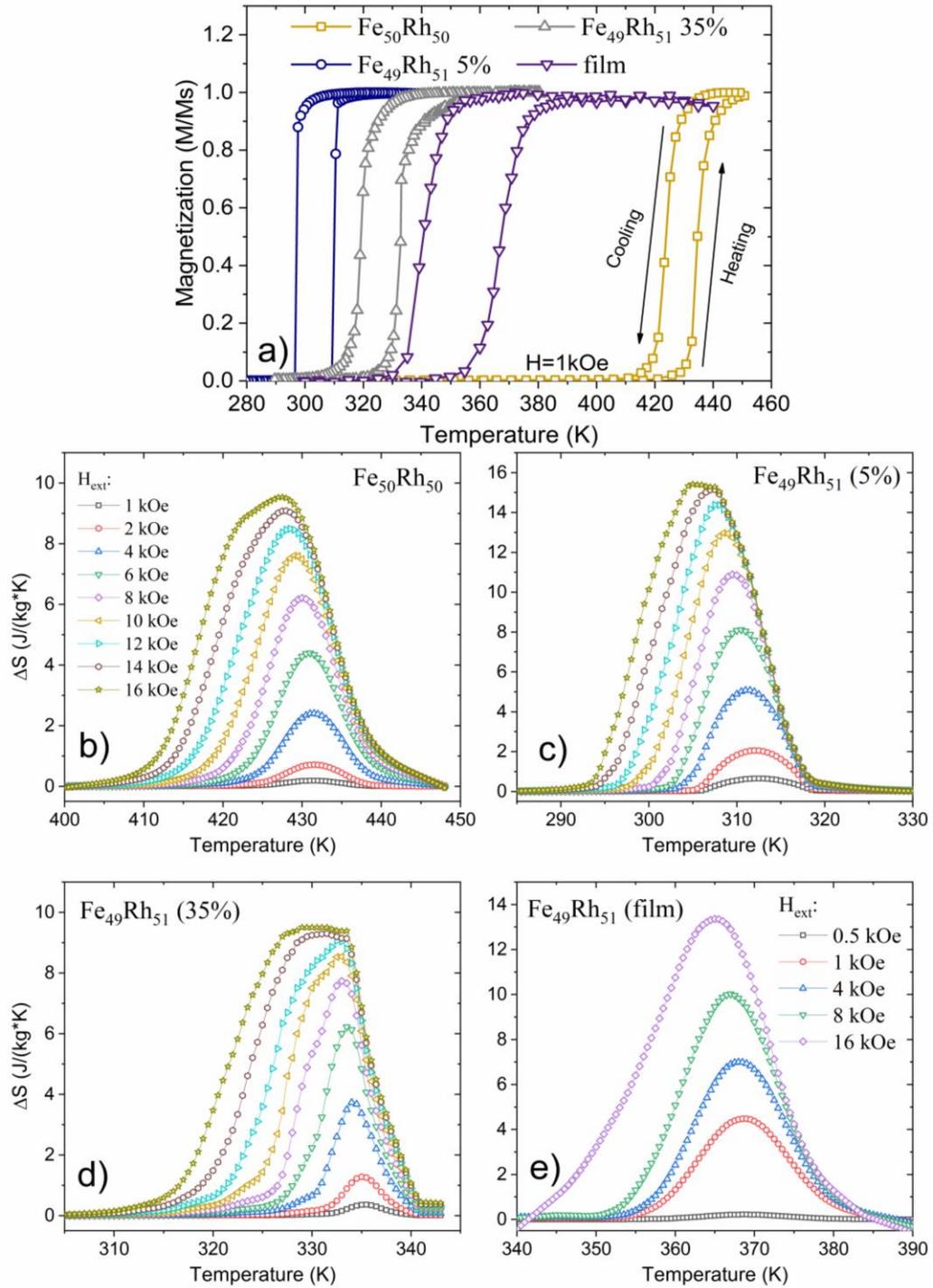

Figure 1. a) The magnetization *vs.* temperature dependences of FeRh samples measured in the field of 1 kOe. The curves are measured using a heating and cooling protocol to determine the temperature hysteresis. b-e) Entropy change for bulk samples and thin films measured in fields up to 16 kOe. The entropy change was estimated from the temperature dependences of the magnetization using the Maxwell relation. The curves in figures b, c, d are obtained for the similar magnetic fields.

The determination of the phase transition temperature for each sample was carried out using vibrating sample magnetometry (VSM). Figure 1a depicts the magnetization temperature dependences in the field of 1 kOe (the field value is given taking into account the demagnetizing factor). The values of phase transition temperatures and temperature hysteresis widths for each of the samples are given in Table 1. The phase transition temperatures of the samples with different elemental compositions are consistent with previously published results [18,47]. The phase transition temperature mismatch for samples with different γ-phase contents could be caused by both mechanical stresses from the γ-phase [21], and a slight deviation in the elemental composition of the $Fe_{49}Rh_{51}$ (35%) alloy from the nominal one [18]. Based on the obtained data, there is a clear tendency for temperature hysteresis in bulk alloys to broad as the γ-phase content increases (Table 1). The influence of the microstructure on the hysteresis properties of the phase transition is expected and described [21,48]. Magnetization temperature dependences in various magnetic fields were measured (**Supplementary materials, Figure S4**). The entropy change was calculated using the Maxwell relation (**Supplementary materials, equations S1, S2**) in order to estimate the caloric properties of the samples (Figure 1b-e). The entropy change calculations in the fields of 1 kOe and 16 kOe are presented in Table 1. It is remarkable that the film sample has an anomalously large magnetocaloric effect in a low magnetic field (1 kOe). The entropy change decreases as the volume fraction of the γ-phase increases. Curie temperatures were determined for all samples (Table 1). The results demonstrate a trend towards an increase in the phase transition temperature with an increase in iron concentration.

Magnetization field dependences (**Supplementary materials, Figure S6**) exhibited different behavior near the saturation field. The temperatures for measuring the field dependences of the magnetization were chosen according to the measurement protocol described in [49]. The magnetization curves were measured at the same temperatures for a more detailed analysis. The effective anisotropy constants were determined (Table 1) using the magnetization curve approximation near the saturation field by the Akulov law [50]. A significant decrease in the film anisotropy constant may be associated with both the influence of external stresses from the side of the single-crystal substrate and a small film thickness [23]. The presence of a paramagnetic γ-phase in the bulk sample leads to pinning of magnetic domain walls at defects. Also, a decrease in the single crystallites size is observed in samples with a γ-phase. These factors can explain the change in the value of effective magnetocrystalline anisotropy.

Table 1. Magnetic and caloric properties of FeRh-based samples: phase transition temperature during heating ($T_{ph.\ tr.}^{\uparrow}$) and cooling ($T_{ph.\ tr.}^{\downarrow}$) in a field of 1 kOe, temperature hysteresis width ($\Delta T$), entropy change ($\Delta S$) in fields of 1 kOe and 16 kOe, effective magnetic anisotropy constant ($K_{eff}$), Curie temperature ($T_C$).

| Sample | $T^{\uparrow}_{ph.\ tr.}$, K (1kOe) | $T^{\downarrow}_{ph.\ tr.}$, K (1kOe) | $\Delta T$, K | $\Delta S$, J/(kg*K) (1kOe) | $\Delta S$, J/(kg*K) (16kOe) | $K_{eff} *10^4$, erg/cc | $T_C$, K |
|---|---|---|---|---|---|---|---|
| $Fe_{50}Rh_{50}$ | 434 | 424 | 10 | 0.2 | 9.5 | 54.7 | 663 |
| $Fe_{49}Rh_{51}$ (5%) | 310 | 297 | 13 | 0.6 | 15.2 | 59.6 | 642 |
| $Fe_{49}Rh_{51}$ (35%) | 333 | 319 | 14 | 0.3 | 11.5 | 124.1 | 651 |
| $Fe_{49}Rh_{51}$ (film) | 367 | 340 | 27 | 4.5 | 13.3 | 6.1 | 640 |

**Magneto-optical Kerr effect measurements**

The hysteresis loops for the film were measured using MOKE (magneto-optical Kerr effect) at different temperatures to analyze the surface magnetic properties. The top film layer is less sensitive to stresses from the substrate. The measurement protocol for the hysteresis loops completely coincided with that described in Ref. [36] in order to correctly compare the results with the VSM data. The temperature dependences of coercivity for two types of measurements are shown in Figure 2a. The results obtained using VSM provide information about the magnetization reversal of the entire sample volume. The hysteresis loops measured with MOKE provide information about the magnetization reversal of the film surface layer (~30 nm) [51]. The increase in coercivity upon cooling is explained by the decrease in size of the ferromagnetic phase to the single-domain state. Another important result is an increase in the phase transition temperature due to the presence of mechanical stresses from the substrate (for more details, see the discussions). Thus, it is necessary to consider the local mechanical stresses in the sample for a correct explanation of the system evolution near the phase transition temperature. It should be emphasized that a lower value of coercivity is observed according to the MOKE results than by VSM. The decrease in coercivity confirms the previously mentioned hypothesis by the authors in [36]. Cooling the film leads to a decrease in the fraction of the ferromagnetic phase near the substrate. Since the magnetic anisotropy of the top layer is smaller, therefore the coercivity of the sample slightly decreases. A comparison of the obtained results with those presented in Ref. [36] enables to identify several stages in the ferromagnetic phase evolution during cooling (Figure 2a,b). The sample surface is in the ferromagnetic state (regions 1 and 2) at temperatures above 350 K. A decrease in temperature leads to a change in the domain structure (Figure 2c). Cooling below 350 K leads to the appearance of an antiferromagnetic phase on the sample surface. The emergence of multidomain ferromagnetic clusters is the reason for the increase in coercivity (region 3). The maximum coercive force is reached near 337 K, which is achieved due to the transition of the ferromagnetic cluster to a single-domain state (Region 4). Most probably, the size distribution of clusters is nonuniform. Therefore, the temperature ranges of the existence of single-domain clusters is approximately determined. A further decrease in temperature (region 5) leads to a decrease in the size of ferromagnetic clusters and their transition to the superparamagnetic state.

The effective domains sizes in various states were estimated using the coercivity temperature dependence results (**Supplementary materials**). The decrease in the domain size during cooling in region 1 is associated with a decrease in the effective thickness of the ferromagnetic phase. An increase in the effective domain size in region 3 may be associated with a decrease in the number of domains caused by a decrease in the size of ferromagnetic clusters. The estimated size of single-domain clusters on the film surface is about 20 nm. This value correlates with other published results [28,52].

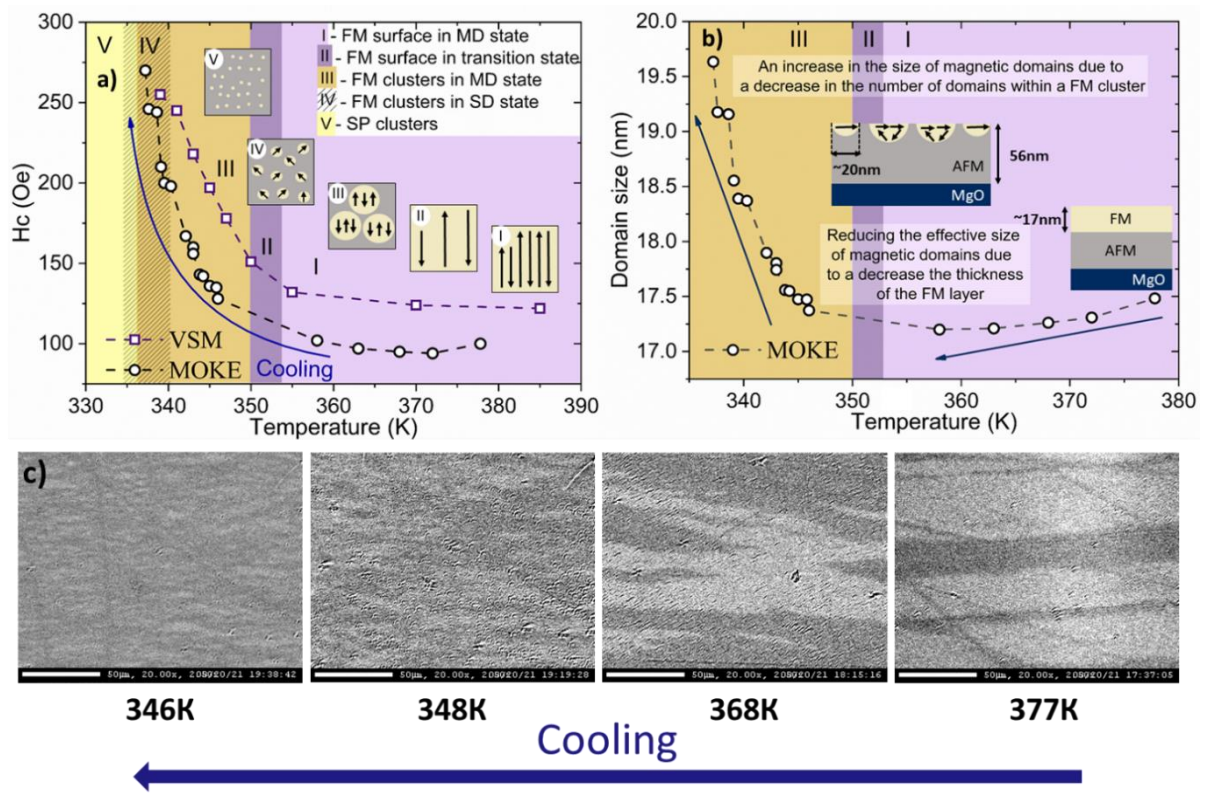

Figure 2. a) Temperature dependence of the coercive force for a film. The measurements were carried out using VSM and MOKE during cooling from the ferromagnetic to the antiferromagnetic state. Different colors indicate the temperature intervals where the sample is in different magnetic states. The type of magnetic state in each region is indicated in the inset. The sketches in the inset schematically illustrate the magnetic state of the sample at different temperatures (top view of the film). The ferromagnetic phase is marked in yellow, the antiferromagnetic phase is marked in gray. The arrows

inside the ferromagnetic phase show the direction of magnetization inside the magnetic domain. b) Temperature dependence of the magnetic domains effective size on the film surface. The effective size of magnetic domains was estimated from the MOKE results. The inserts in the picture schematically illustrate the magnetic state of the film (side view) in the low-temperature and high-temperature regions. c) MOKE pictures of the film surface taken at different temperatures (during cooling).

**Temperature magnetic force measurements**

MFM (magnetic force measurements) studies were carried out on a two-phase $Fe_{49}Rh_{51}$ (35%) sample at various temperatures in order to study the ferromagnetic phase nucleation processes in detail. The sample phase transition temperature is within the region accessible for experimentation. The $Fe_{49}Rh_{51}$ (35%) phase transition is more controlled (less abrupt) in temperature compared to the $Fe_{49}Rh_{51}$ (5%) sample. Therefore, it is possible to study the features of the ferromagnetic phase formation near α/γ-phase interface in a two-phase sample. Figure 3a shows MFM images of the same area of the sample taken at different temperatures during the heating process. The γ-phase localization is marked to aid in the interpretation of the results. The size and shape of the γ-phase regions according to the MFM images results coincide with the results obtained using The MFM image results show that the size and shape of the γ-phase regions coincide with the electron microscope results (Figure 3c).

The sample is in an antiferromagnetic state (region I in Figure 3b) at a temperature of 303 K. The presence of insignificant stray magnetic fields (residual ferromagnetic phase) near the scratch at this temperature is noticeable. Further heating up to a temperature of 322 K is accompanied by the nucleation of ferromagnetic clusters on local surface defects (region II in Figure 3b). It is worth noting that the bright yellow regions are interpreted as ferromagnetic clusters, because they have the largest stray magnetic fields due to their size and topology. It is also important to note that ferromagnetic clusters appear in places farthest from the γ-phase localization. A sharp increase in the proportion of the ferromagnetic phase on the sample surface during heating from 322 K to 324 K occurs due to the processes of its nucleation and growth (region III in Figure 3b). A further increase in the magnetization upon heating to 332 K is provided by the growth and nucleation of ferromagnetic clusters on the sample surface (region IV in Figure 3b). The merging of ferromagnetic clusters leads to a reducing of the magnetic stray fields, which is noticeable in Figure 3a. It is also worth noting that the processes occurring in regions III and IV lead to an increase in magnetization, according to the VSM results (Figure 3b). The sharp increase in the magnetization (in the region V) may be associated with an increase in the ferromagnetic phase fraction in the volume of the sample. The decrease in stray fields in the MFM results confirms the validity of this hypothesis. Heating the sample above the temperature of 336 K is accompanied by the merging of ferromagnetic clusters, which remained localized near the the γ-phase interface (region VI). Further significant heating above the temperature of 360 K transforms the sample into a fully ferromagnetic state. It should also be noted that the $Fe_{49}Rh_{51}$ (5%) sample has a significantly lower content of the γ-phase. Therefore, the transition region VI is not noticeable in Figure 1a. The presence of a domain structure inside the ferromagnetic phase can be traced on the MFM image at a temperature of 352K. The sizes and shapes of the domains are in qualitative agreement with the resultsin [53].

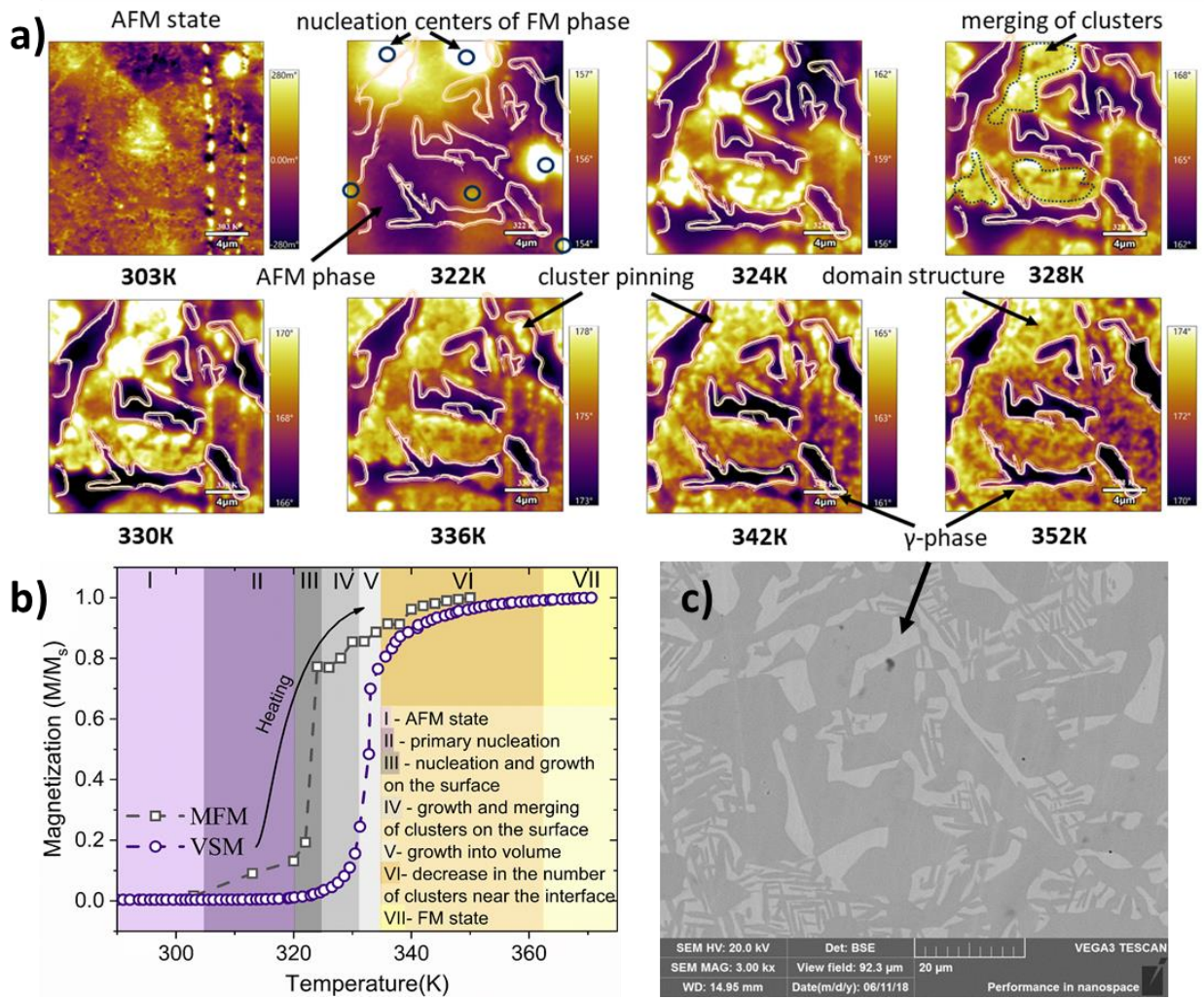

Figure 3. a) MFM images of the $Fe_{49}Rh_{51}$ (35%) sample surface taken at different temperatures (during heating). The pictures were taken from the same area. The paramagnetic γ-phase localization sites are denoted by light lines. b) Magnetization temperature dependences of a two-phase sample obtained by the MFM and VSM methods (during heating). The colors indicate the temperature intervals in which various mechanisms of the ferromagnetic phase evolution predominate. The ferromagnetic phase evolution mechanisms for each region are specified in the inset. c) SEM image of the $Fe_{49}Rh_{51}$ (35%) sample surface. The main (α) phase is marked in dark gray, the paramagnetic γ-phase is indicated in light gray. The investigated areas of the sample on the SEM and MFM images are not the same.

We found a correlation between the phase transition mechanisms described above in the two-phase bulk alloys and the mechanisms dominating in thin films.

**Magnetization relaxation measurements**

We studied the dynamic behavior of the ferromagnetic phase evolution process to refine the proposed model. The protocol for measuring the magnetization time dependences near the phase transition temperature is detailed in Ref. [36]. Figures 4a,b depict the most typical magnetization time dependences for the samples. Figure 4a shows the relaxation curves measured in the state where the ferromagnetic phase grows into the sample volume (region V in Figure. 4b). Figure 4b shows relaxation curves for samples in state VI. As can be seen from the experimental curves, the high signal-to-noise ratio of the film sample does not allow us to further study the behavior of this system in more detail. It should be noted that the phase transition in the $Fe_{49}Rh_{51}$ (5%) sample occurs abruptly (Figure 1a). For this reason, it was not possible to measure the magnetization relaxation dependences using VSM (**Supplementary materials, Figure S10**). Figure 4a shows the presence of characteristic steps in the magnetization time dependences for all samples. The deviation of the growth law from the exponential one may be explained by the presence of the constraint and the magnetocaloric effects (more details in the discussions). It is also worth noting that the presence of the γ-phase causes the shape of the experimental curves closer to exponential ones. For a more detailed analysis of the phase transition evolution, we estimated the characteristic relaxation times at different temperatures for a more detailed analysis of the phase transition evolution in different samples (Figure 4c, the time estimation process is described in the **Supplementary materials**). The temperature dependences of the relaxation time parameters are presented in normalized form (relative to the phase transition temperature) for ease of comparison. For all samples the characteristic forms of the obtained dependences are the same. The general tendency towards a decrease in the relaxation time with increasing temperature is due to an increase in the volume fraction of the ferromagnetic phase [49]. The anomalous behavior near the phase transition temperature may be associated with the intense growth of the ferromagnetic phase causing the appearance of mechanical stresses. The relaxation time of the two-phase sample is longer than that of the single-phase one at low temperatures. This behavior is explained by the predominance of the nucleation processes in this temperature range. We believe that the two-phase sample has a large number of defects; therefore, it has a larger number of nucleation centers. The processes of the new phase growth begin to predominate approaching the phase transition temperature. The inclusion of the γ-phase in a two-phase sample limits the growth and merging of ferromagnetic clusters. A decrease in the contribution of these processes leads to a decrease in the relaxation time in the two-phase sample.

In addition, a detailed analysis of the stepwise behavior of the experimental curves was carried out. For this purpose, the position of the steps (**Supplementary materials, Figure S11**) was selected on each curve. Its characteristic parameters (the magnitude of the magnetization jump, its time duration and the magnitude of the magnetization in this state) were determined for each step. It was noted above that it is necessary to take into account the mechanical stresses in the system for the correct interpretation of the results. Therefore, the estimation of the step parameters for the magnetization relaxation dependences was done by an analogy with the the fatigue cracks growth analysis. The fatigue crack growth occurs due to the accumulation of stresses near the defect. The crack length change under the load action obeys a power law depending on the stress intensity factor according to the Paris equation [54]. By an analogy, we established the relationship between $\frac{dM}{dt}$ and $M_n$ (the magnetization of the sample at the moment the nth step occurs). As a result, the analysis of temporal relaxations at different temperatures was described $\frac{dM}{dt} \sim A * (M_n)^n$ (Figure 4d) (the reasons for choosing such a dependence are described in the **Discussion** and **Supplementary materials**). The value of the exponent *n* was determined by approximation. It should also be noted that the dependences $\frac{dM}{dt}(M_n)$ exhibit nonmonotonic behavior. The observed peak is associated with a change in the mechanism of ferromagnetic phase evolution at different states (for more details, see the **Discussions**). The temperature dependences of the parameter *n* were obtained (Figure 4e) as a result of the steps analysis on the relaxation curves for the $Fe_{50}Rh_{50}$ and $Fe_{49}Rh_{51}$ (35%) samples. It is interesting that the obtained nonmonotonic curves are extremums close to the temperatures of mechanisms of ferromagnetic phase evolution change. For example, at temperatures up to 326 K, nucleation processes predominate in the two-phase sample (region III in Figure 3b), so the parameter n has small value. The occurrence of cluster merger (accompanied by an increase in mechanical stresses) leads to an increase in the exponent *n*. The accumulated stresses relax due to the intense ferromagnetic phase growth (the γ-phase serves as a buffer for mechanical stresses) as the temperature increases. This is accompanied by a decrease in the exponent *n*. The second example of nonmonotonic behavior for the two-phase sample is observed at the temperature of 337 K (the beginning of region VI). There are no nucleation processes and no intense ferromagnetic phase growth in this temperature range. The slow relaxation of mechanical stresses at the cluster merging state leads to an increase in the exponent *n*.

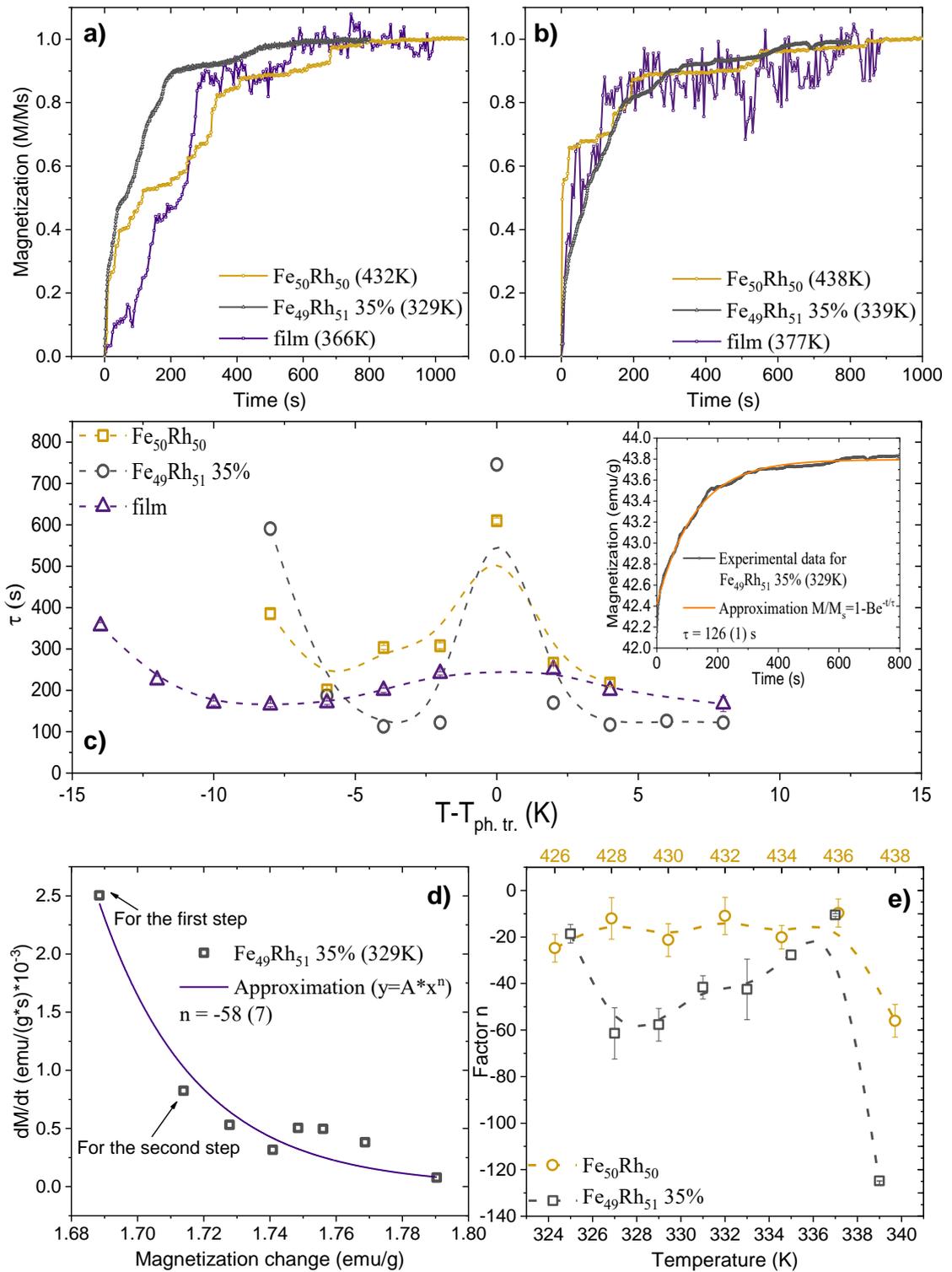

Figure 4. Magnetization time dependences of $Fe_{50}Rh_{50}$, $Fe_{49}Rh_{51}$ (35%) and film samples at a fixed temperature and external magnetic field (1 kOe) in a) a low-temperature region and b) a high-temperature region near the phase transition temperature. The high noise/signal ratio for the film sample is due to the low value of the magnetic moment. c) Temperature dependence of relaxation time. The temperature scale is normalized to the temperature of the phase transition for each of the samples. The external magnetic field is 1 kOe. The inset shows the approximate time dependence of the magnetization, from which the relaxation time was determined.

d) The dependence of the magnetization jump ratio to the step duration time is a function of the sample magnetization in this state. The presented dependence was obtained by analyzing the relaxation curve at the temperature of 329K for the $Fe_{49}Rh_{51}$ (35%) sample. The experimental results are approximated by a power law. e) Temperature dependences of the exponent n for $Fe_{50}Rh_{50}$ and $Fe_{49}Rh_{51}$ (35%) samples. The temperature scale for $Fe_{50}Rh_{50}$ is indicated at the top. The results for the film sample are from [36].

**Modeling and computational calculations**

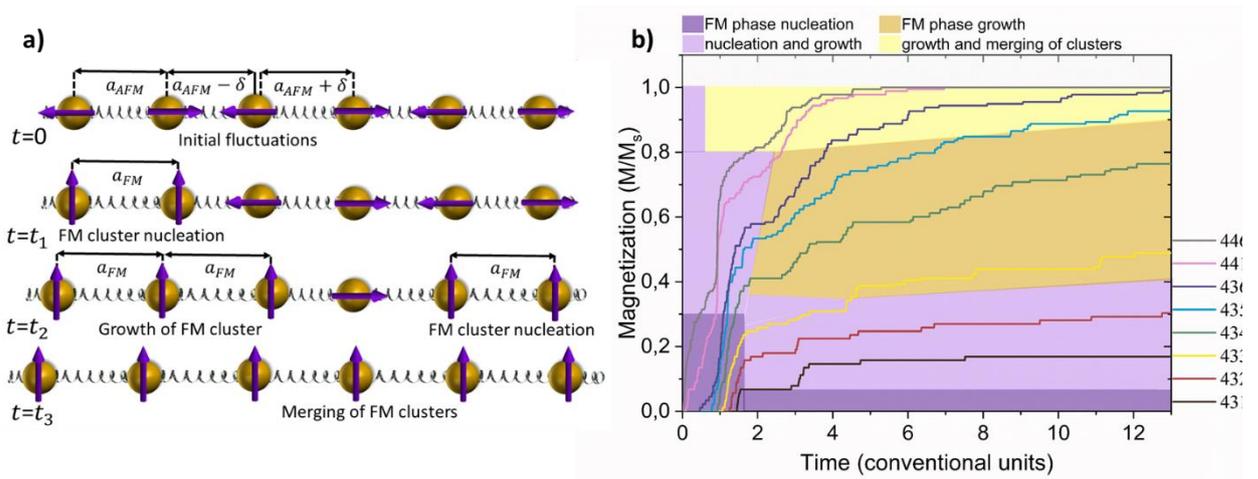

Figure 5. a) Schematic illustration of an elastically bound N-atom chain at various states in time. At the initial moment of time, the system was in an antiferromagnetic state. Atoms are moved out from their equilibrium position. The processes of nucleation, growth, and merging of ferromagnetic clusters are schematically presented below. The purple arrows indicate the direction of the magnetic moments. b) Magnetization time dependence of the system for various initial conditions. Different colors mark the regions where different mechanisms of ferromagnetic phase evolution predominate.

The phase transition process from the antiferromagnetic to the ferromagnetic state was simulated by the molecular dynamics method in order to illustrate the hypothesis of the correlation between long-term magnetization relaxation and mechanical parameters. The model considers a one-dimensional chain of N-atom elastically interacting with each other (Figure 5a). The distance between atoms at the initial moment of time is equal to $a \pm \delta$. Where $a$ is the lattice constant in the antiferromagnetic state, $\delta$ is the atoms's deviation from the equilibrium position. The parameter $\delta$ corresponds to a lognormal distribution, the most probable value of which is proportional to temperature. The condition for the transition of atoms from the antiferromagnetic to the ferromagnetic state is an increase in the distance between neighboring atoms up to the lattice constant value in the ferromagnetic state(for more details about the model, see the **Supplementary**

**materials**). The proposed model makes it possible to analyze the magnetic phase transition temporal behavior by taking into account the emerging local stresses between atoms. The results of modeling the magnetization time dependence in an atom chain for various initial conditions are shown in Figure 5b. The obtained curves demonstrate good qualitative agreement with the experimental dependences (Figure 4a,b). This result is particularly surprising because our model relies on a very strong and rather speculative approximation. The dominant processes in the phase evolution at different times and at different temperatures during the relaxation process were identified (Figure 5b). Analysis shows that the most considerableand fast jumps in magnetization (at the beginning of the relaxation process) are associated with the nucleation of the ferromagnetic phase. The size of the emerging ferromagnetic clusters depends on the temperature. An increase in temperature leads both to an increase in the average size of emerging clusters and the emergence rate of new clusters. Also, an increase in temperature leads to the complete indistinguishability of the nucleation and phase growth processes at the initial moment of time are almost inseparable. They are present at the same time. The delay in the growth of magnetization in the middle of the relaxation process are related to the appearance of stress in the regions between two ferromagnetic clusters, which relax along the chain of atoms. It takes some time for a fluctuation of the atom position to occur in this region leading to a transition to the ferromagnetic state. The final part of the relaxation process is associated both with nucleation and growth of the ferromagnetic phase at low temperatures. An increase in temperature leads to the fact that the final part of the relaxation is mainly determined by the growth of the ferromagnetic phase. The growth of the phase is accompanied by the processes of ferromagnetic clusters merging at temperatures at temperatures of almost completely transformation to the ferromagnetic state. It should be noted that according to the simulation results, the merging of ferromagnetic clusters occurs only at the final stage of the phase transition. The one-dimensional model revealed no regularity, indicating that the ferromagnetic phase nucleation would form in priority from the chain's edge. It is possible that the three-dimensional model of the phase transition will demonstrate that the phase nucleation occurs from the sample's surface, as experimental works [22,28] have shown.

## Discussion

A comprehensive analysis of the VSM, MFM and MOKE results allows us to establish the main stages of the ferromagnetic phase formation at various temperatures. Nucleation and growth of the ferromagnetic phase begin at the surface of the sample. The merging of ferromagnetic clusters dominates after the formation of the ferromagnetic phase inside the sample volume. It is also interesting to note that the ferromagnetic phase nucleation begins far away from the γ-phase. This result can be explained from the point of view of mechanical stresses arising near the γ-phase. The presence of tensile stresses near the α/γ-phase interface increases the phase transition temperature [21]. As a result, residual antiferromagnetic regions are localized near the γ-phase interface. Consequently, the γ-phase's presence, size and percentage influence the formation of the dominant ferromagnetic phase evolution mechanisms. A change in the dominant factors, as a result, leads to a change in the shape of magnetization temperature dependence (Figure 1a). Also, it was possible to determine the micromagnetic state of ferromagnetic clusters at different temperatures by analyzing the discussed experimental data. The decrease in the number of magnetic domains correlates with the decrease in the size of the ferromagnetic cluster. However, these results characterize only the stationary states of the samples.

The dynamic behavior of the ferromagnetic phase evolution near the phase transition temperature is of the greatest interest. The magnetization time dependences of the samples demonstrate a number of unexpected features. Firstly, the relaxation time is anomalously long for a ferromagnet. Secondly, steps in the magnetization time dependences are observed under fixed external conditions. To explain the observed features, it is necessary to consider possible mechanisms that can cause a deceleration of the ferromagnetic phase growth.

The α-phase transition temperature of a binary alloy from an antiferromagnetic to a ferromagnetic state depends on the mechanical stresses. An increase in the lattice constant during the antiferromagnetic-ferromagnetic transition leads to the appearance of local mechanical stresses. As a consequence, the phase transition temperature in this region increases, which slows down the phase growth. This effect is called the constraint effect in the literature[55]. Mechanical stresses, on the other hand, can relax in a solid. The growth of the ferromagnetic phase continues after the relaxation of mechanical stresses. It follows from the discussed results that the phase transition temperature increases for samples with the initial mechanical stresses ($Fe_{49}Rh_{51}$ 35%, film). The presence of an additional γ-phase in the bulk alloy ensures the appearance of external stresses at the interface between the phases. Mechanical stresses are generated due to the influence of the substrate on the thin film. Magneto-optical measurements, which demonstrate a decrease in the phase transition temperature (for the "stress-free" film layer) in the near-surface film layer, verify this conclusion.

The second mechanism leading to a decrease in the ferromagnetic phase growth rate is the magnetocaloric effect. The growth phase is accompanied by a decrease in temperature. The sample is locally cooled due to heat transfer processes that slows down the phase growth. The phase growth in this place continues again after the part of the sample pass into thermal equilibrium. The entropy change estimates show that this effect is significant for all samples. The film sample has a large magnetocaloric effect in a low magnetic field. This agrees with the published results[56].

Therefore, it can be assumed that this mechanism of the phase growth stopping dominates in film, rather than the constraint-effect.

Changing the sample microstructure can alter the dominant processes of phase evolution during a phase transition. The presence of a large amount of the γ-phase in the sample, for example, spatially limits the ferromagnetic phase growth. Therefore, ferromagnetic phase nucleation is the predominant mechanism in such a system. Furthermore, the γ-phase acts as a buffer for the mechanical stresses relaxation. The presence of this feature contributes to a decrease in the relaxation time (compared to a single-phase sample) at temperatures when the phase growth processes dominate (Figure 4c). Figure 4e clearly shows an increase in the parameter $n$ for the two-phase sample compared to a single-phase sample at the beginning of the phase transition. An increase in this value is associated with a change in the dominant mechanisms of the phase evolution. It should be noted that the problem of creating micro- and nanostructured materials is relevant. For example, the use of the methods for structuring FeRh alloys described in [44–46], will make it possible to create well-ordered two-phase structures.

There is currently no first-principle description of the observed relaxation phenomena. However, the results of the proposed phenomenological model, which combines the phonon (associated with a change in the lattice constant) and magnon (associated with a change in the magnetic substructure) properties of the FeRh alloy, are correlated with the experiment. The agreement between the model and experimental results allows us to separate the evolution phase mechanisms in time. It should be noted that the model can be improved by adding rhodium atoms. This should introduce anharmonicity into the system, which may lead to the discovery of new effects. It is also necessary to change the function describing the interaction between atoms, which in general is not linear and will differ in the antiferromagnetic and ferromagnetic states. However, more important is the question of the phase transition occurrence criteria. Modern methods for calculating phase transitions within the framework of an effective field [14,57,58] do not predict phase nucleation processes. Although this process is the key one for first-order phase transitions. Other theoretical methods [59,60] define the processes of phase formation *a priori*. Although the proposed phenomenological phase transition criterion describes the nucleation process well, a more rigorous approach will be required in the future.

## Conclusion

An analysis of the presented results in this work allows us, for the first time (according to the information found), to establish the features of the main ferromagnetic phase evolution mechanisms near the temperature of the first-order phase transition. The following processes were distinguished: (I) primary nucleation, (II) nucleation and growth at the surface, (III) growth and merging of the ferromagnetic clusters, (IV) pinning of the antiferromagnetic phase near the γ-phase localization. Moreover, the stages of the ferromagnetic phase growing on the surface and in the volume of the bulk sample were separated. The proposed methods open up the prospect for an indirect estimation of the ferromagnetic phase evolution inside the sample volume. These approaches may be beneficial for studies of first-order phase transitions.

It was shown that it is possible to tune the dominant mechanisms of the ferromagnetic phase evolution by changing the microstructure of the sample. Thereby, the two-phase materials exhibit shorter magnetization relaxation times. It is expected that controlled nanostructuring will be able to increase the speed of spintronic elements and magnetocaloric devices. We demonstrated a new approach to study dynamic processes through experimental and model consideration of phase evolution time behavior. These techniques give the possibility of separating the ferromagnetic phase nucleation, growth, and merging processes in time. The authors believe that a thorough investigation of the ferromagnetic phase evolution mechanisms will be beneficial from both a fundamental and applied standpoint.

Methods:

**Synthesis**

Bulk samples of the nominal compositions $Fe_{50}Rh_{50}$ and $Fe_{49}Rh_{51}$ were synthesized by arc melting of the pure elements Fe (99.995), Rh (99.98) in an argon atmosphere in a water-cooled Cu-crucible. The samples were then sealed in quartz tubes and annealed for 72 h at 1273 K, followed by water quenching. For the experiment, the samples were cut into the forms of long thin plates by a wire saw with minimal destructive effects. In order to carry out microscopic investigations, the surface was prepared using standard metallographic techniques: grinding on SiC paper and polishing with diamond suspension and Al and Si oxide slurries.

The film was grown on a single-crystalline MgO (001) substrate at 773 K by co-evaporation of Fe and Rh in an ultrahigh-vacuum molecular beam epitaxy chamber. The base pressure was maintained at ~$10^{-10}$ Torr. Before the growth, the substrates were annealed at 873 K for 1 h to obtain flat surfaces. After the growth, the film was post-annealed for 1.5 h at 893 K More detailed information on the synthesis and structural data of the film sample can be found in [36].

**Structure data**

XRD analysis was performed on the Bruker D8 Discover. The elemental composition and homogeneity of the resulting phase were verified by SEM Tescan Vega 3 with the EDX option. The Adobe Photoshop was used to analyze the images.

**VSM**

Magnetic measurements were studied using the VSM Lakeshore 7407 Series over a wide temperature range (80–800 K). Heating and cooling of the sample were carried out at a rate of 2 K/min, overheating did not exceed 0.5 K. The magnetization field dependences were measured from the demagnetized state using the protocol is described below. The sample was transferred from the antiferromagnetic to the ferromagnetic state by heating in a zero magnetic field. After that, the magnetization curve was measured. Heating the sample from the antiferromagnetic to the ferromagnetic state in a zero magnetic field ensured zero residual magnetization.

**MOKE**

The Evico Magnetics MOKE microscope was used for measuring hysteresis loops and imaging magnetic domain structure. Due to the weak contrast in the longitudinal mode, all images were obtained by subtracting a background image recorded at a fully magnetized state. To perform temperature-controlled measurements with the MOKE microscope, a custom-made thermostat was used. The sample was placed on an aluminum holder connected to a heating system. At the center of the holder, a Pt1000 temperature sensor was mounted at the same level as the sample. The sensor was connected to a Keithley 2400 source meter by the four-probe method. The sensor was calibrated in MPMS XL 7 to reduce the uncertainty of the temperature measurement. The sample holder and the heater were placed in a thermostat with a transparent viewing window. Due to the limited space, a 20x objective lens with a relatively large focusing distance combined with an additional 2.5x lens was used. Before measuring a hysteresis loop, the sample was heated to T = 380 K and then slowly cooled down to the target temperature. The temperature instability did not exceed 0.2 K.

**MFM**

Magnetic force microscopy (MFM) images were obtained in a commercial scanning probe microscope MFP-3D (Asylum Research, Goleta, CA, USA) using the MFM01 (Tipsnano, Tallinn, Estonia) probe coating CoCr with curvature radius ~ 40nm and the spring constant of 3 N/m. The magnetic field ($B_{ext}$ = 0.8 kOe) was applied perpendicular to the plane of the sample.

The evolution of the domain structure was studied using a heating stage with the temperature range from 296 to 573 K (the temperature was stabilized at 0.05 K). MFM images were obtained in the temperature range of 303-352 K with the step of 2 K.

MFM is a dual-path tapping mode technique that uses a magnetic tip for the scans. During the first scan, the system operates as per the normal tapping mode, recording the topography of the sample. During the second scan, the tip is lifted from the sample surface by a fixed height (usually a few tens or hundreds of nanometers, in our case 150 nm). It then retraces the surface profile recorded in the first scan while maintaining the lift height.

The long-range force interactions (i.e., force gradients) between the magnetic probe and the magnetic sample in MFM are recorded and correlated in the second pass from the shift in frequency ($\Delta\omega$), amplitude ($\Delta A$), or phase ($\Delta\phi$) from the initial driven parameters (i.e., $\omega_0$, $A_0$, and $\phi_0$, respectively) of the oscillating cantilever [61].

**Magnetization relaxation measurements**

To exclude the influence of temperature hysteresis in Fe-Rh alloys, which has a noticeable effect on the results reproducibility, time-dependent magnetic measurements were carried out so that the phase transition occurs in the same direction. A measurement protocol, described below, was developed to provide this requirement. The sample was cooled to pure antiferromagnetic state in a zero magnetic field before each measurement. Then, it was heated up to the required temperature (overheating no more than 0.5 K). The sample was kept for 5 min to ensure the system temperature equilibrium. After that, the magnetic field was turned on. Measurements of the time dependence of magnetization were begun after magnetic field stabilization. This measurement technique provides us with reproducibility of results and avoids the demagnetization procedure required in the case of a pure FM state as a starting point. Otherwise, after the third stage, it would be necessary to demagnetize the sample to eliminate the residual magnetization from the previous measurement. Upon cooling to the AFM state, the residual magnetization becomes zero. Besides, we exclude from the experimental data the records made during the magnetic field ramping process.

**Computational calculations**

The system dynamics was calculated using the C++ program. The lognormal distribution was given by the lognormal_distribution<> function. The deviations of atoms from the equilibrium position were set using the <random> library. The time step was chosen to ensure the convergence of the calculated results. Convergence was determined by calculation at various time steps. The number of atoms in the chain was 180.

**Data availability**

All relevant data are available from the authors.

**Competing interests**

The authors declare no competing financial interests.

**Acknowledgment**

The authors are grateful to Dr. Yuye Wu (Beihang University) and Olga Shuleshova (Leibniz Institute for Solid State and Materials Research Dresden) for sample preparation. The research was carried out with the financial support of the Russian Science Foundation within the framework of project 22-22-00291. The work was supported in part by the M.V. Lomonosov MSU Program of Development. Komlev A.S. thanks the BASIS Foundation for scholarship support. T.S.I. and D.A.K. are grateful the Ministry of Education and Science of the Russian Federation (project № 075-15-2021-696) for using equipment MFM measurement which were performed of the Joint Research Center «Material Science and Metallurgy» NUST MISIS.

**Author contributions**: **A.S.K.** designed the work concept and measurements protocols, performed static and dynamic magnetic measurements, theoretical model and computational calculations, processed experimental data, and prepared the first draft of the manuscript with contributions from all other authors; **R.A.M.** repeated magnetic measurements in order to verify them, processed part of the electron microscopy measurement data; **T.S.I.** and **D.A.K.** carried out MFM measurements; **A.M.C.** synthesized and prepared samples, performed entropy change calculations, participated in editing the text of the manuscript; **N.A.K.** and **A.S.V.** carried out MOKE measurements, participated in editing the text of the manuscript; **V.I.Z.** coordinated the project, participated in editing the text of the manuscript; **N.S.P.** performed formal analysis, provided supervision, participated in editing the text of the manuscript. All the authors commented on successive drafts and have given their approval to the final version of the manuscript.

References

1.	Nikitin, S. A. *et al.* The magnetocaloric effect in Fe49Rh51 compound. *Physics Letters, Section A: General, Atomic and Solid State Physics* **148**, 363–366 (1990).

2.	Chirkova, A. *et al.* Giant adiabatic temperature change in FeRh alloys evidenced by direct measurements under cyclic conditions. *Acta Materialia* **106**, 15–21 (2016).

3.	Baranov, N. V., Zemlyanski, S. V. & Kamenev, K. Electrical Resistivity and Phase Transitions in FeRh Based Compounds: Influence of Spin Fluctuations. in *Itinerant Electron Magnetism: Fluctuation*

*Effects* (eds. Wagner, D., Brauneck, W. & Solontsov, A.) 345–351 (Springer Netherlands, 1998). doi:10.1007/978-94-011-5080-4_21.

4. Ibarra, M. R. & Algarabel, P. A. Giant volume magnetostriction in the FeRh alloy. *Phys. Rev. B* **50**, 4196–4199 (1994).

5. Aubert, A. *et al.* Simultaneous Multi-Property Probing During Magneto-Structural Phase Transitions: An Element-Specific and Macroscopic Hysteresis Characterization at ID12 of the ESRF. *IEEE Transactions on Instrumentation and Measurement* **71**, 1–9 (2022).

6. Modak, R. *et al.* Phase-transition-induced giant Thomson effect for thermoelectric cooling. *Applied Physics Reviews* **9**, 011414 (2022).

7. Jiménez, M. J., Schvval, A. B. & Cabeza, G. F. Ab initio study of FeRh alloy properties. *Computational Materials Science* **172**, 109385 (2020).

8. Gottschall, T. *et al.* Making a Cool Choice: The Materials Library of Magnetic Refrigeration. *Advanced Energy Materials* **9**, 1901322 (2019).

9. Wu, H. *et al.* Current-induced Néel order switching facilitated by magnetic phase transition. *Nat Commun* **13**, 1629 (2022).

10. Waske, A. *et al.* Energy harvesting near room temperature using a thermomagnetic generator with a pretzel-like magnetic flux topology. *Nat Energy* **4**, 68–74 (2019).

11. Komlev, A. S. & Zverev, V. I. Chapter 14 - Magnetocaloric effect for medical applications. in *Magnetic Materials and Technologies for Medical Applications* (ed. Tishin, A. M.) 437–467 (Woodhead Publishing, 2022). doi:10.1016/B978-0-12-822532-5.00001-7.

12. Komlev, A. S., Gimaev, R. R. & Zverev, V. I. Smart magnetocaloric coatings for implants: Controlled drug release for targeted delivery. *Physics Open* **7**, 100063 (2021).

13. Thiele, J.-U., Maat, S. & Fullerton, E. E. FeRh/FePt exchange spring films for thermally assisted magnetic recording media. *Applied Physics Letters* **82**, 2859–2861 (2003).

14. Derlet, P. M. Landau-Heisenberg Hamiltonian model for FeRh. *Phys. Rev. B* **85**, 174431 (2012).

15. Ultrafast Emergence of Ferromagnetism in Antiferromagnetic FeRh in High Magnetic Fields. https://scholar.google.com/citations?view_op=view_citation&hl=ru&user=PgCAsSYAAAAJ&sortby=pubdate&citation_for_view=PgCAsSYAAAAJ:Bg7qf7VwUHIC.

16. Pressacco, F. *et al.* Subpicosecond metamagnetic phase transition in FeRh driven by non-equilibrium electron dynamics. *Nat Commun* **12**, 5088 (2021).

17. Lewis, L. H., Marrows, C. H. & Langridge, S. Coupled magnetic, structural, and electronic phase transitions in FeRh. *J. Phys. D: Appl. Phys.* **49**, 323002 (2016).

18. Gimaev, R. R., Vaulin, A. A., Gubkin, A. F. & Zverev, V. I. Peculiarities of Magnetic and Magnetocaloric Properties of Fe–Rh Alloys in the Range of Antiferromagnet–Ferromagnet Transition. *Physics of Metals and Metallography* **121**, (2020).

19. L. I. Vinokurova, A. V. Vlasov, & M. Pardavi-Horváth. Pressure effects on magnetic phase transitions in FeRh and FeRhIr alloys. *physica status solidi (b)* **78**, 353–357 (1976).

20. Saha, P. *et al.* Effect of substrate and Fe/Rh stoichiometry on first order antiferromagnetic–ferromagnetic transition in FeRh thin films. *Journal of Magnetism and Magnetic Materials* **551**, 169095 (2022).

21. Chirkova, A. *et al.* The effect of the microstructure on the antiferromagnetic to ferromagnetic transition in FeRh alloys. *Acta Materialia* **131**, 31–38 (2017).

22. Kumar, H. *et al.* Strain effects on the magnetic order of epitaxial FeRh thin films. *Journal of Applied Physics* **124**, 085306 (2018).

23. Arregi, J. A., Caha, O. & Uhlíř, V. Evolution of strain across the magnetostructural phase transition in epitaxial FeRh films on different substrates. *Phys. Rev. B* **101**, 174413 (2020).

24. Warren, J. L., Barton, C. W., Bull, C. & Thomson, T. Topography dependence of the metamagnetic phase transition in FeRh thin films. *Sci Rep* **10**, 4030 (2020).

25. Lee, Y. *et al.* Large resistivity modulation in mixed-phase metallic systems. *Nat Commun* **6**, 5959 (2015).

26. Keavney, D. J. *et al.* Phase Coexistence and Kinetic Arrest in the Magnetostructural Transition of the Ordered Alloy FeRh. *Scientific Reports* **8**, 1–7 (2018).

27. Arregi, J. A. *et al.* Magnetization reversal and confinement effects across the metamagnetic phase transition in mesoscale FeRh structures. *J. Phys. D: Appl. Phys.* **51**, 105001 (2018).

28. Almeida, T. P. *et al.* Direct visualization of the magnetostructural phase transition in nanoscale FeRh thin films using differential phase contrast imaging. *Phys. Rev. Materials* **4**, 034410 (2020).

29. Gatel, C. *et al.* Inhomogeneous spatial distribution of the magnetic transition in an iron-rhodium thin film. *Nat Commun* **8**, 15703 (2017).

30. Bergman, B. *et al.* Identifying growth mechanisms for laser-induced magnetization in FeRh. *Phys. Rev. B* **73**, 060407 (2006).

31. Awari, N. *et al.* Monitoring laser-induced magnetization in FeRh by transient terahertz emission spectroscopy. *Appl. Phys. Lett.* **117**, 122407 (2020).

32. Li, G. *et al.* Ultrafast kinetics of the antiferromagnetic-ferromagnetic phase transition in FeRh. *Nature Communications* **13**, (2022).

33. Frazer, T. D. *et al.* Optical transient grating pumped X-ray diffraction microscopy for studying mesoscale structural dynamics. *Sci Rep* **11**, 19322 (2021).

34. Mariager, S. O. *et al.* Structural and Magnetic Dynamics of a Laser Induced Phase Transition in FeRh. *Phys. Rev. Lett.* **108**, 087201 (2012).

35. Radu, I. *et al.* Laser-induced generation and quenching of magnetization on FeRh studied with time-resolved x-ray magnetic circular dichroism. *Phys. Rev. B* **81**, 104415 (2010).

36. Komlev, A. S. *et al.* Ferromagnetic phase nucleation and its growth evolution in FeRh thin films. *Journal of Alloys and Compounds* **874**, 159924 (2021).

37. Manekar, M. & Roy, S. B. Nucleation and growth dynamics across the antiferromagnetic to ferromagnetic transition in (Fe0.975Ni0.025)50Rh50: analogy with crystallization. *J. Phys.: Condens. Matter* **20**, 325208 (2008).

38. Manekar, M., Chattopadhyay, M. K. & Roy, S. B. Glassy dynamics in magnetization across the first order ferromagnetic to antiferromagnetic transition in Fe0.955Ni0.045Rh. *J. Phys.: Condens. Matter* **23**, 086001 (2011).

39. Lu, W., Nam, N. T. & Suzuki, T. Magnetic Properties and Phase Transition Kinetics of $Fe_{50}(Rh_{1-x}Pt_{x})_{50}$ Thin Films. *IEEE Transactions on Magnetics* **45**, 4011–4014 (2009).

40. Massey, J. R. *et al.* Asymmetric magnetic relaxation behavior of domains and domain walls observed through the FeRh first-order metamagnetic phase transition. *Phys. Rev. B* **102**, 144304 (2020).

41. Massey, J. R. *et al.* Phase boundary exchange coupling in the mixed magnetic phase regime of a Pd-doped FeRh epilayer. *Phys. Rev. Materials* **4**, 024403 (2020).

42. Wang, Y. *et al.* Spin pumping during the antiferromagnetic–ferromagnetic phase transition of iron–rhodium. *Nature Communications* **11**, 275 (2020).

43. Nan, T. *et al.* Electric-field control of spin dynamics during magnetic phase transitions. *Science Advances* **6**, eabd2613.

44. Merkel, D. G. *et al.* A Three-Dimensional Analysis of Magnetic Nanopattern Formation in FeRh Thin Films on MgO Substrates: Implications for Spintronic Devices. *ACS Appl. Nano Mater.* (2022) doi:10.1021/acsanm.2c00511.

45. Ye, X. *et al.* Creating a Ferromagnetic Ground State with Tc Above Room Temperature in a Paramagnetic Alloy through Non-Equilibrium Nanostructuring. *Advanced Materials* **34**, 2108793 (2022).

46. Nadarajah, R. *et al.* Towards laser printing of magnetocaloric structures by inducing a magnetic phase transition in iron-rhodium nanoparticles. *Sci Rep* **11**, 13719 (2021).

47. Suzuki, I., Naito, T., Itoh, M. & Taniyama, T. Barkhausen-like antiferromagnetic to ferromagnetic phase transition driven by spin polarized current. *Appl. Phys. Lett.* **107**, 082408 (2015).

48. Scheibel, F. *et al.* Hysteresis Design of Magnetocaloric Materials—From Basic Mechanisms to Applications. *Energy Technology* **6**, 1397–1428 (2018).

49. Komlev, A. S. *et al.* Correlation between magnetic and crystal structural sublattices in palladium-doped FeRh alloys: Analysis of the metamagnetic phase transition driving forces. *Journal of Alloys and Compounds* **898**, 163092 (2022).

50. Akulov N.S. Uber den Verlauf der Magnetisierungskurve in starken Feldern. *Zeitschrift fur Phys.* **69**, (1931).

51. Soldatov, I. V. & Schäfer, R. Selective sensitivity in Kerr microscopy. *Review of Scientific Instruments* **88**, 073701 (2017).

52. Grimes, M. *et al.* X-ray investigation of long-range antiferromagnetic ordering in FeRh. *AIP Advances* **12**, 035048 (2022).

53. Yokoyama, Y. *et al.* MFM observation of magnetic phase transitions in ordered FeRh systems. *Journal of Magnetism and Magnetic Materials* **177–181**, 181–182 (1998).

54. Parton, V. Z. *Fracture mechanics: from theory to practice*. (CRC Press, 1992).

55. Karpenkov, D. Yu. *et al.* Pressure Dependence of Magnetic Properties in ${\mathrm{La}(\mathrm{Fe},\mathrm{Si})}_{13}$: Multistimulus Responsiveness of Caloric Effects by Modeling and Experiment. *Phys. Rev. Applied* **13**, 034014 (2020).

56. Manipulation of magnetocaloric effect in FeRh films by epitaxial growth - ScienceDirect. https://www.sciencedirect.com/science/article/pii/S0925838822009653?via%3Dihub.

57. Sandratskii, L. M. & Buczek, P. Lifetimes and chirality of spin waves in antiferromagnetic and ferromagnetic FeRh from the perspective of time-dependent density functional theory. *Phys. Rev. B* **85**, 020406 (2012).

58. Phys. Rev. B 101, 014444 (2020) - Self-consistent local mean-field theory for phase transitions and magnetic properties of FeRh. https://journals.aps.org/prb/abstract/10.1103/PhysRevB.101.014444.

59. De Bruijn, T. J. W., De Jong, W. A. & Van Den Berg, P. J. Kinetic parameters in Avrami—Erofeev type reactions from isothermal and non-isothermal experiments. *Thermochimica Acta* **45**, 315–325 (1981).

60. Kolmogorov, A. N. Izv Akad Nauk SSSR. *Ser Materm* **3**, 355–359 (1937).

61. Kazakova, O. *et al.* Frontiers of magnetic force microscopy. *Journal of Applied Physics* **125**, 060901 (2019).

# Highlights

- The main stages of the ferromagnetic phase evolution during the phase transition are distinguished.
- The ferromagnetic phase nucleation regions and the features of phase growth are determined.
- The mechanisms that lead to a decrease in the phase growth rate are established
- The phase transition kinetics in samples with different microstructures are studied.